# An Implementation of Web Services for Inter-Connectivity of Information Systems

**Aftab Ahmed Chandio[1, 2, 3], Dingju Zhu[1], Ali Hassan Sodhro[1, 2], and Muhammad Umer Syed[1, 2]**

*[1]Shenzhen Institutes of Advanced Technology Chinese Academy of Sciences, Shenzhen, China*
*[2]Graduate University of Chinese Academy of Sciences, Beijing, China*
*[3]Institute of Mathematics and Computer Science University of Sindh, Jamshoro, Pakistan*

**Abstract:** As educational institutions and their departments rapidly increase, a communication between their end-users becomes more and more difficult in traditional online management systems (OMS). However, the end-users, i.e., employees, teaching staff, and students are associated to different sub-domains and using different sub-systems that are executed on different platforms following different administrative policies. Because of their intercommunication is not automated integrated, consequently, the overall efficiency of the system is degraded and the communication time is increased. Therefore, a technique for better interoperability and automated integration of departments is an urgent needed. Many of existing systems does not have a set of connections yet, such as the system of the University of Sindh (UoS). In this paper, we propose a system for the UoS, named integration of inter-connectivity of information system (i3) based on service oriented architecture (SOA) with web services. The system i3 monitors and exchanges the student's information in support of verification along heterogeneous and decentralized nature. Moreover, the proposed system provides capability of interoperability between their sub-systems that are deployed in different departments of UoS and using different programming languages and database management systems (DBMS).

## 1. INTRODUCTION

Educational institutions and their departments are rapidly growing that increase a huge number of end-users (i.e., students, teaching staff, and employees). For instance, a system for a university is enclosed between several departments' sub-system that are frequently communicated. Such sub-systems include the system of: (a) extended campuses, (b) library, (c) student, and (d) administration information management departments [1] [2]. In the aforementioned systems their end-users are strongly correlated with each other for their frequent communications (e.g., information exchanging and sharing). Due to the strong limitations, many of these systems presently provide a manual communication. Such limitations are: (1) a system in each department of institution works on a different platform (i.e., operating system) using (2) different programming languages (i.e., Java, and Dot.Net) and (3) different database management systems (DBMS) (i.e., SQL Server, MySQL, and Oracle).

Due to the heterogeneity of platforms, languages, and DBMS in their sub-systems, a manual communication makes a tough and time consumed way for exchanging information. Because of the aforementioned statement, the organizations are forced to enhance the overall existed systems in order to achieve quality of services (QoS) (i.e., fast response time). Specifically, the existing systems do not create an automatic interconnection between their departments at the time of end-users of departments need to exchange information. In spite the above fact, the delay in communication and the response time subsequently increases. Therefore, the departments in the institution are needed to integrate with each other in order to communicate automatically. Their results must facilitate the several departments for information exchanging and sharing.

The stimulation of new and rapid technological changes i.e., service oriented architecture (SOA) with heterogeneous and decentralized manners, can enable the organizations to frequent upgrades and changes of the systems. For the aforementioned issues, we propose a system for a university that is based on SOA architecture. SOA can facilitate with a group of Web Services (WSs) over the network. Furthermore, SOA also provides the way for making an enterprise system with distributed computers, databases, and connections. Specifically, the group of WSs is the heart of SOA architecture, which plays a pivotal role in SOA architecture and provides business logic as a service for the heterogeneous environment [11]. By the help of WSs it could be possible to extend the previous system as a new system integrating their departments without replacing of the overall deployed system [3]. WS is a technique used for quickness and security, ability of interoperability, integration, and platform- and language-independent in the system.

This paper proposes a system integration of inter-connectivity of information system (i3) for the University of Sindh[1] (UoS) Pakistan. The system i3 [16] solves the problem that frequently occurs in the process of no-dues verification of a student from different departments. This process is needed at the time of final certificate issuing to the student by higher and confidential authority of the university. In the system i3, WSs are used to realize the SOA behavior, wherein different departments in the UoS automatically integrate and exchange information. Moreover, this paper characterizes different levels of information which is being exchanged between different departments in the university. In this paper, we briefly address five different modules used in the proposed system. The application of each module is implemented in different programming languages. Their data is stored in different DBMS.

The rest of the paper is organized as follows. Section 2 discusses the state-of-the-art followed by Section 3 that states a basic overview of the proposed system. In the Section 4, this paper explains experimental work and evolution of the system. In the final, we present the discussion in Section 5 and conclude the paper in Section 6.

## 2.    STATE-OF-THE-ART

As large number of organizations is deploying their applications on the Internet, SOA is one of the emerging architecture for distributing and integrating multiple services. The system in SOA works in distributed computing environment. Tanenbaum [4] precised the distributed computing such that the distributed system is "a collection of independent computers that appears to its users as a single coherent system as the aspect deal with software". Specifically, the basic nature of the distributed computing is a "single program as a service can be run on network between multiple computers simultaneously".

In 1980s, program-to-program was a fashionable work in the concept of distributed computing approach. Several companies had been utilized this approach. Such variants of the above approach are: (a) system application architecture (SAA) by IBM working on own LU 6.2 API and PU 2.1 connection protocol, (b) remote procedure calls (RPCs) by Apollo, (c) distributed network architecture (DNA) by Microsoft, (d) CORBA by object management group (OMG), and (e) DCE by the open software foundation. The aforementioned approaches suffer with strong limitations, i.e., the data format, API, and network environment with hard-programming and limited application types. Alternatively, WS is a distributed computing approach that supports easy programming to find automatically distributed services (i.e., request and response). Moreover, it enables to the system to process a communication in different APIs [5]. WS is a technological backbone of SOA that provides the way of design, development, deployment, and management of business logic (i.e. functions, messages, and services) over the Internet. The major goal of SOA services is to make scalable, flexible, and loosely-coupled services to exchange information [6].

WS life cycle is comprised of three major roles: (a) client, (b) broker, and (c) server. They are explained as follow and shown in Fig. 1.

- **Client:** A client sends a request to a broker in order to access the services.

- **Broker:** A broker performs an intermediate role between client and server. It has service registry/directory to identify the service.

- **Server:** A server is a service provider to the client.

The complete life cycle of WS works with four basic elements, i.e., (a) Simple Object Access Protocol (SOAP), (b) eXtensible Markup Language (XML), (c) Universal Description, Discovery and Integration (UDDI), and (d) Web Service Description Language (WSDL). The services (i.e., requests and response) are circulated according to the above three roles in the form of XML, which use SOAP protocol for transferring the messages. SOAP protocol consists of three natures, i.e., *find*, *publish*, and *bind* used in a complete process in the WS life cycle [7]. Basically, SOAP protocol is the XML-based protocol that can be used for applications to exchange information over HTTP Internet protocol. Moreover, SOAP protocol also provides mechanisms for the distributed environment, including, cross platform independency, language independency, and interoperability. UDDI is a directory for storing WSDL

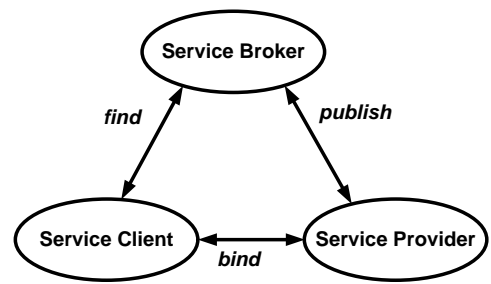

Figure 1.    Major roles in Web Services

---

information, while WSDL is used to self-describe and locate the WSs [12] [13].

A vast body of research has been taken advantage of SOA and WS for solving real-world problems (such as, discussed in [2], [5], [7] - [11]). For instances, Alkhanak [2] addressed the SOAPGS system that has been created for the University of Malaysia. The system facilitates to postgraduate students with many services for reducing cost and time in their daily academic life when students need to communicate to other departments. Panian [3] expressed benefits of integration in an enterprise system. The author explained that the integration process gives forward-looking and flexible way where organizations can easily and quickly view data on the spot, which performs operations in real-time. Li [7] studied on the hardware platform with the help of web services. The author described some addition code is needed in both, server-side and client-side of Java mobile phone. Liang [8] discussed about some factors of the university system that need to integrate between their applications. Author described several factors can effect to management of universities. Such factors are as a shortage of funds and some technical problems including the departments in the university independently working, the lake of standard planning, different departments are based on different levels of information technology, "Information isolated island" prominent, and decision making support system [8]. A. Shaikh [9] discussed that the web services is the solution of increasing problems in distributed telemedicine system during data integration, vendor lock-in, and interoperability. Furthermore, Talis SOA architecture [11] provided Talis adaptor which contains a group of services support heterogeneous behavior for different sources.

## 3. PROPOSED MODEL

The proposed system i3 is designed to share and exchange the information that is associated with students of different departments in the institution. The system i3 facilitates each department easily and quickly to find the status of student for verification purpose. Our proposed model is viable solution of the problems that frequently occurred in the university information system of UoS. Specifically, the verification process that occurs at the time of issuing certificate is needed by the student and university administration. Higher authority of the university issues the students' degree certificate after passing the required examination for the degree program. Before the issuing degree certificate, each student has to submit the no-dues certificate stamped by related departments in the university. This process increases the: (a) overall response time and (b) workloads on employees. The system i3 makes it possible that the higher authority of the university can easily and quickly get a status report of the student from all departments at first hand.

In the deployed system of UoS, the applications in the departments are totally differed then others applications in the departments. Basically, the applications in each department implemented in different programming tools and database systems. To integrate all of the departments of the university in order to exchange information is a solution of without replacing deployed system. However, due to their heterogeneity nature, the integration process is a big challenge in the above system.

As the main goal of our proposed model is sharing and exchanging information into different departments, in this section, we characterize information that use in the entire system. Characterization is the process which helps in understanding the overall behavior of the system highlighting [17] [18]. In our proposed system, the information in each department can be characterized into five basic levels of information. Fig. 2 shows their relationships among different levels of information system and their description are shown in Table I.

TABLE I. INFORMATION LEVELS

| No. | Levels of information | Descriptions |
|-----|----------------------|--------------|
| 1 | Admission Level (AL) | About student' information |
| 2 | Hostel Level (HL) | About hostel' and student' information |
| 3 | Library Level (LL) | About library's and student' information |
| 4 | Campus Level (CL) | About student' information |
| 5 | Examination Level (EL) | About student's examination |

The information in admission level (AL) is a basic and primary student's information, i.e., personal data and educational information of student. A personal data of student includes student id, first name, last name, address, and contact number. The educational information of student consists of institution name, department name, degree program, and graduation year. The AL information is accessed by all of the departments as clearly shown in the figure.

Information in two other levels, hostel level (HL) and library level (LL), accessed by the examination department of the university. Specifically, LL information includes books' information (i.e., book id, isbn, title, author, publisher, and year) and students' information (i.e., personal and educational student information). On the next, HL information includes rooms' and students' information. It is noted that the both levels of information (i.e., LL and HL) provide a special report. Such report shows a list of defaulter students, i.e., the students has been: (a) issued books and (b) allotted the room. As the university provides different degree programs, the university can extend the academic extensions in time to time, called campus in different cities. We call information of campuses level (CL) to information used in academic extensions located in the distance. CL in each campus stores the student information (i.e., personal and educational information). Final, information in examination level (EL) is very important and secure information, which manages the examination records. EL uses the students' examination records for issuing final certificates. In consequence, the secret information is needed to exchange among different departments for the verification purpose.

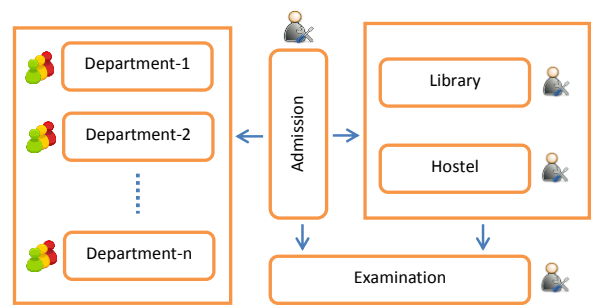

Figure 2. Basic Levels in the System i3 for information exchanging

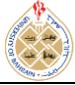

Without system i3, these departments are not able to automatically exchange data rather they only communicate manually. For example, all departments accesses similar primary student's information, while the examination department uses information from multiple departments including LL and HL. Integration is the best solution for such type of enterprise system. The integration of knowledge and information process will assist in bringing up some standards for information collection, dissemination and management, and also some other standards cataloguing, storing and retrieval of library data [1].

## 4. EXPERIMENTAL WORK

### A. Evaluation

The system i3 in UoS deploys several WSs without any costly or hard programming technique. The system i3 has five modules which create five information systems (i.e., already summarized in the previous section). In this Section, we describe all modules in brief. Each module is deployed to perform a specific function and used different platforms realizing the nature of heterogeneity. In order to make automatic connectivity, few modules provide the WSs to integrate the other modules. Such modules are: (1) the administration management information system (AMIS), (2) the library management information system (LMIS), (3) the hostel management information system (HMIS), (4) the campus information system (Campus IS), and (5) the examination management information system (EMIS).

### 1) AMIS

AMIS plays a fundamental role in the system i3, this module stores primary students' information (i.e., personal and educational information). Since all modules need to use a basic student's information for the registration purpose, AMIS module provides the services for sharing such information, called WSs. The WSs in AMIS is used by LMIS, HMIS, Campus IS, and EMIS. For example, the Fig. 3 shows a student registration page of the LMIS module which is accessing WSs of AMIS for registering the student in own system. In the left portion of the figure, the student's list is shown, which is provided by AMIS. Java programming technique and the Oracle database has been used in AMIS module.

### 2) LMIS

This module is used in the university library to manage the book's information, called LL information. It also accesses WSs of AMIS for registering the student's information in the system. Furthermore, LMIS module maintains the status of each student such that books are issued to students, and provides the list of defaulter students. The status of each student is needed by EMIS and accessed as in a defaulter student information report that will be used by EMIS as WSs. LMIS module has also been implemented in Java programming together SQL server 2005 to store their information.

### 3) HMIS

Similar to LMIS, HMIS also accesses service from AMIS for a student's registration in own database. It maintains the status of all rooms in the hostel and the students' information, called HL information. It provides a special information report of a room allotted to student. The report and the status of room and student observed as WSs, which is only required by EMIS module. This module has been implemented in Java for programming and MySQL server for database.

### 4) Campus IS

The number of the Campus's information system may increase in the future for a new extension of the university. This module also use AMIS service for student registration in own system. DotNet technology is used in programming for its application and Microsoft Access database is used for storing database.

### 5) EMIS

EMIS works in the last phase when students want to get certificates from higher authority of university after completing theirs degree program. In this stage, EMIS's application needs to access all services along special reports from multiple departments at the same time. The certificate page of EMIS is shown in Fig. 4.

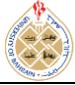

Figure 3. The Student Registration Page in LMIS accessing Web Services of AMIS

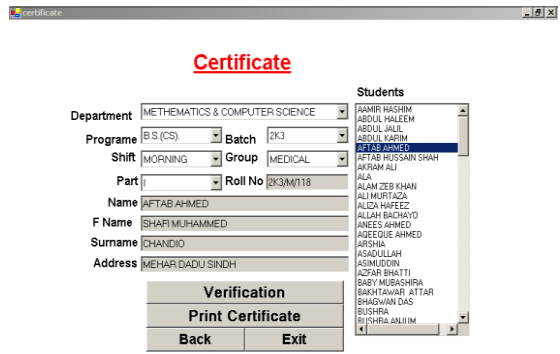

Figure 4. The Certificate Page in EMIS accessing Web Services of (AMIS, HMIS, and LMIS)

The left portion of the figure shows a student list taken from AMIS with the help of WSs. After selecting a student's information, the authority person can click the verification button in order to find a student's status. Then the student's status will be verified from several departments, which uses WSs provided by LMIS and HMIS. The application of this module has been created in DotNet technology, while the database used in Oracle.

### B. Setup

System i3 in UoS has been developed based on the project of Apache software foundation for web services implementation called Apache eXtensible Interaction System (Axis). Axis is an open source SOAP engine [14]. We have implemented SOAP in the Axis server, whereas all the services are created in Java programming but it also can be programmed with C/C++.

When we deploy any service in the Axis engine, we need to use of deployment descriptor, which is pure XML-based code, called web service deployment descriptor (WSDD) [15]. In WSDD, we write several nodes, such as (a) the deployment node for describing services in service name node, (b) the parameter name node for theirs methods and functions accessed by clients, and (c) the beanMapping node for describing a bean of their service. Axis un-deploys any service by using un-deployment node. In the below WSDD XML-based File 1, we have created three services for deployment in the system i3 with their methods and functions. The services and theirs descriptions are shown in Table II.

TABLE II.    WS SERVICES OF THE i3

| No. | Services | Descriptions |
|---|---|---|
| 1 | *AdmissionDataBaseManager* () | Accessed by LMIS, HMIS, Campus IS, and EMIS |
| 2 | *LibraryDataBaseManager* () | Accessed by EMIS |
| 3 | *HostelDataBaseManager* () | Accessed by EMIS |

All services provide a collection of different methods and related records. The first service accesses the admission level information. It consists of several methods/functions and four different bean records, i.e., student record, department record, department record, and program record. The second service belongs to library level information and consists of a bean record, i.e., library student record. While in the last service, the hostel level information serves clients. The first service is used by all modules in the system i3, while remaining both services, second and last, are only accessed by the examination module for the verification purpose.

## 5.    DISCUSSION

In order to discuss our proposed system i3, we pilot its complete life cycle, shown in Fig. 5. The figure shows that web service approach is making a self-describable service and combining internal and external services to realize SOA architecture.

The service requesters (i.e., Campuses, LMIS, HMIS, EMIS) send a request to service provider (i.e., AMIS services) to access a student's information. The service requester, EMIS needs to access services from multiple service providers (i.e., LMIS, HMIS and AMIS) at the same time in order to monitor the student's status for the verification purpose. Initially, in the WSs environment, the service requesters find the location of service, which is published by service provider and registered in discovery service. After finding service location, all service requesters communicate service location to receive WSDL file that provides a list of all services and the way of usage. Then clients can make a connection to the location of service and invoke functions for exchanging information through the XML based message in SOAP protocol.

System i3 can allow similar and dissimilar applications to communicate and exchange data each

```
<deployment xmlns="http://xml.apache.org/axis/wsdd/"
              xmlns:java="http://xml.apache.org/axis/wsdd/providers/java" >
<handler name="print" type="java:LogHandler"/>
<service name="AdmissionDataBaseManagerService" provider="java:RPC">
      <requestFlow><handler type="print"/></requestFlow>
      <beanMapping qname="myNS:StudentRecord" xmlns:myNS="urn:BeanService"
              languageSpecificType="java:StudentRecord"/>
      <beanMapping qname="myNS:DepartmentRecord" xmlns:myNS="urn:BeanService"
              languageSpecificType="java:DepartmentRecord"/>
      <beanMapping qname="myNS:ProgrammeRecord" xmlns:myNS="urn:BeanService"
              languageSpecificType="java:ProgrammeRecord"/>
      <beanMapping qname="myNS:ListItem" xmlns:myNS="urn:BeanService"
              languageSpecificType="java:ListItem"/>
</service>
<service name="LibraryDataBaseManagerService" provider="java:RPC">
      <requestFlow> <handler type="print"/> </requestFlow>
      <beanMapping qname="myNS:LibraryStudentRecord" xmlns:myNS="urn:BeanService"
              languageSpecificType="java:LibraryStudentRecord"/>
</service>
<service name="HostelDataBaseManagerService" provider="java:RPC">
      <requestFlow> <handler type="print"/> </requestFlow>
      <beanMapping qname="myNS:HostelStudentRecord" xmlns:myNS="urn:BeanService"
              languageSpecificType="java:HostelStudentRecord"/>
</service>
</deployment>
```

File 1.    WSDD XML-based

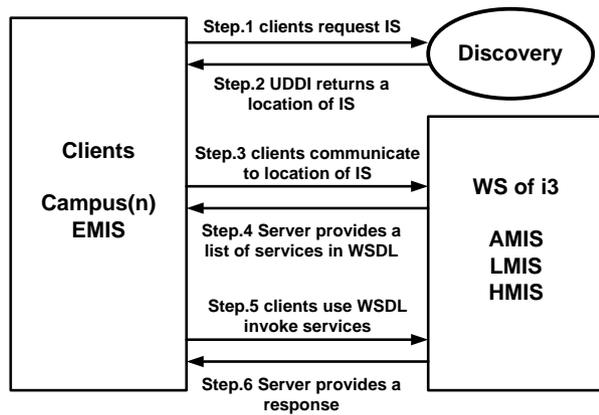

Figure 5.    The Life-cycle of the System i3 using the Web Services

other through the XML based message in SOAP protocol. We have deployed above three services in the Apache Axis SOAP engine. It allows each modules of the system i3, which can work on the platform independent as well as language independent, even store information on different databases. For example, in the system i3 LMIS works in Java platform with the Microsoft SQL server database, and AMIS in Java platform with Oracle database, Java services from both modules are accessed in EMIS module developed in dotNet platform.

## 6.    CONCLUSION

In this paper, we have solved the real-world problem of university and proposed system i3. In the system i3, we use web services, which is the best solution for SOA infrastructure. We integrated different systems of University of Sindh (UoS) in one framework, since without the system i3 they could not work automatically for frequent intercommunicating. System i3 solved a major problem of the examination department in UoS. The problem was raised during verification of student from all departments for issuing a final certificate. The problem is more complicate because of applications in all departments function on different types of platforms and databases. Our proposed system i3 cannot be only implemented in UoS, but also can be fit in other institutions facing similar problems.

### ACKNOWLEDGMENT

The shorter version [16] of the paper has been published in the proceedings of the IMECS'12 at Hong Kong, China on March 2012.

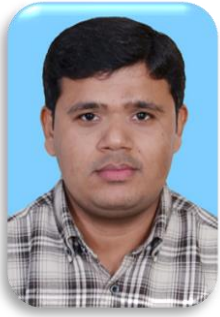

**Aftab Ahmed Chandio** is a doctoral student at Shenzhen Institutes of Advanced Technology, Chinese Academy of Sciences, Shenzhen, Guangdong, China. He is also a lecturer of Institute of Mathematics and Computer Science at the University of Sindh, Jamshoro, Sindh, Pakistan. His research interests include cloud computing, parallel and distributed systems, scheduling, workload characterization, and map matching strategies for GPS trajectories.

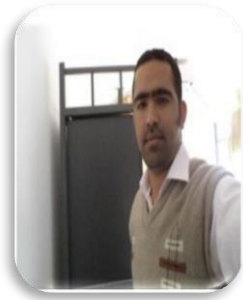

**Ali Hassan Sodhro** is a doctoral student at Shenzhen Institutes of Advanced Technology, Chinese Academy of Sciences, Shenzhen, Guangdong, China. His research interests include transmission power control, energy optimization in wireless body area networks (WBANs), medical QoS, and medical video transmission in WBANs.

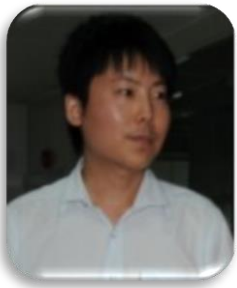

**Dingju Zhu** received the Ph.D. degree from Institute of Computing Technology, Chinese Academy of Sciences, China. He is an associate professor of Shenzhen Institutes of Advanced Technology, Chinese Academy of Sciences, Shenzhen, Guangdong, China. His research interests include cloud computing and parallel and distributed computing.

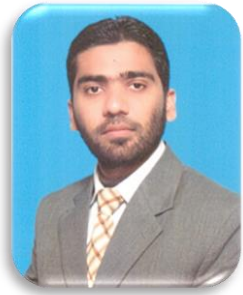

**Muhammad Umer Syed** is working towards a master degree program at Shenzhen Institutes of Advanced Technology, Chinese Academy of Sciences, Shenzhen, Guangdong, China. His research interests include embedded systems, surveillance, robot navigation, localization, mobile robot, intelligent systems, docking, obstacle avoidance, auto-recharging, home security, low-cost, and novelty.